# Rare-Earth Ion Coupling Implements Attention-Like Reservoir Computing

Junyan Chen[1], Xinzhe Li[1], Jinsong Fu[1], Axin Du[1], Jinfeng Yao[5], Shuang Gao[1], Wenzhao Sun[6], Limin Jin[1,*], Can Huang[1,3,5,*], Qinghai Song[1,2,3,4,5,*]

[1] Ministry of Industry and Information Technology Key Lab of Micro-Nano Optoelectronic Information System, Guangdong Provincial Key Laboratory of Semiconductor Optoelectronic Materials and Intelligent Photonic Systems, State Key Laboratory of Robotics and Systems (HIT), Harbin Institute of Technology, Shenzhen, 518055, China.

[2] Pengcheng Laboratory, Shenzhen 518055, China.

[3] Quantum Science Center of Guangdong-Hongkong Macao Greater Bay Area, Shenzhen 518055, China.

[4] Collaborative Innovation Center of Extreme Optics, Shanxi University, Taiyuan 030006, Shanxi, China.

[5] Heilongjiang Provincial Key Laboratory of Advanced Quantum Functional Materials and Sensor Devices, Harbin Institute of Technology, Harbin 150001, China.

[6] City University of Hong Kong (Dongguan), Dongguan 523808, China

*Corresponding author: huangcan@hit.edu.cn; jinlimin@hit.edu.cn; qinghai.song@hit.educ.cn

## Abstract:

We present a physical computing paradigm that harnesses the intrinsic nonlinear dynamics of rare-earth-doped core-shell nanoparticles ($Tm^{3+}$@$Er^{3+}$) as a computational substrate. By directly exploiting cross-relaxation and energy-transfer upconversion processes, the system realizes a state-dependent transfer function whose effective decay rate evolves with the instantaneous $Er^{3+}$ population, which mathematically analogous to gating and attention mechanisms in recurrent neural networks. The three spectrally resolved emission channels inherently span disparate timescales, endowing the reservoir with native multi-timescale feature extraction without auxiliary engineering. Under the reservoir computing framework, the coupled three-channel system achieves a total memory capacity exceeding fourfold that of a single-ion reservoir; capacity decomposition further reveals that the nonzero cross-memory capacity is a direct signature of many-body $Tm^{3+}$@$Er^{3+}$ coupling. On the Mackey–Glass and Santa Fe chaotic benchmarks, the system attains normalized mean squared errors of $1.2 \times 10^{-3}$ and $2.1 \times 10^{-2}$, respectively, with only 125 virtual nodes. These results establish rare-earth nanoparticles as a compelling platform for compact and hardware-integrable neuromorphic computing, and introduce "inward evolution", the deliberate exploitation of intra-material quantum dynamics, as a generalizable design principle for next-generation physical computing systems.

## Introduction

Time-series modeling—spanning weather forecasting, electroencephalogram (EEG) analysis, speech recognition, and financial prediction—represents one of the most fundamental and demanding challenges in artificial intelligence. Recurrent neural networks (RNNs) architectures including long short-term memory (LSTM), liquid time-constant networks (LTC), and Transformers have achieved substantial advances on complex temporal tasks by incorporating state-dependent dynamic update mechanisms: gating units that regulate information flow and attention modules that selectively reweight historical context [1–4]. Yet these performance gains exact a steep computational cost; the training and inference of such models impose prohibitive energy demands, severely limiting their deployment on edge devices, mobile platforms, and space-constrained systems [5–7].

Optical neural networks (ONNs) and, more broadly, physical neuromorphic computing, which exploit the parallelism of optical propagation and the intrinsic dynamics of physical substrates to process information, have emerged as a compelling route to circumvent these bottlenecks [8–10]. Nevertheless, most existing optical computing architectures share a fundamental constraint: their evolution operators are linear and time-invariant, their transfer functions are rigidly determined by fixed material constants, and genuine nonlinear responses are absent [11]. Overcoming this limitation has conventionally required the introduction of external delay lines, spatial node arrays, or elaborate feedback loops [12–14]. Although such "outward-scaling" strategies have demonstrated laboratory-level efficacy, they inevitably impose precision optical alignment requirements, hinder miniaturization, and introduce additional analog-to-digital conversion overhead, contradicting the foundational objective of compact, low-power physical computing.

Here we pursue a fundamentally different strategy: rather than augmenting the state space through external auxiliary structures, we directly mine computational resources from the microscopic dynamics intrinsic to the material itself [24]. In rare-earth-doped core-shell nanoparticles ($Tm^{3+}$@$Er^{3+}$), many-body interactions between $Tm^{3+}$ and $Er^{3+}$ ions—specifically cross-relaxation and energy transfer upconversion—cause the effective decay rate of $Tm^{3+}$ to vary dynamically with the instantaneous population of $Er^{3+}$ [15–17]. Consequently, the system's transfer function is no longer a static constant but evolves continuously with the internal state, a behavior that is mathematically analogous to the state-dependent gating and input-dependent attention weighting central to recurrent neural architectures. Concurrently, phonon-assisted relaxation and the selective routing of energy-transfer pathways give rise to three spectrally distinct emission channels with markedly heterogeneous decay times, endowing the system with intrinsic multi-timescale feature extraction capability. Critically, properties such as state-dependent dynamics and multi-scale parallel response, emerge entirely from the material's inherent many-body interactions, requiring no external delay lines, feedback loops, or spatial multiplexing.

To rigorously quantify the information-processing capability of this mechanism, we employ the reservoir computing framework [18–20]. The $Tm^{3+}$@$Er^{3+}$ coupled system achieves a memory capacity of 3.1, more than fourfold that of the uncoupled single-$Tm^{3+}$ reservoir. Capacity decomposition further reveals that the nonzero cross-memory capacity contribution is a direct and experimentally measurable fingerprint of the inter-ion coupling, providing information-theoretic evidence of state-dependent dynamics. On canonical benchmark tasks, the system attains an NMSE of $1.2 \times 10^{-3}$ on the Mackey-Glass chaotic sequence and $2.1 \times 10^{-2}$ on the Santa Fe chaotic sequence, surpassing the majority of reported physical computing systems [23, 25–31]. Because the dynamical resources are encoded in the material's intrinsic multilevel structure, and the nanoparticles can be synthesized via scalable solution-phase methods and deposited as thin films through spin-coating, the platform is inherently compatible with CMOS integration processes. These attributes collectively establish rare-earth-doped nanoparticles as a promising foundation for compact, low-power, and integrable neuromorphic hardware suited to edge and space-constrained deployment scenarios.

## Results

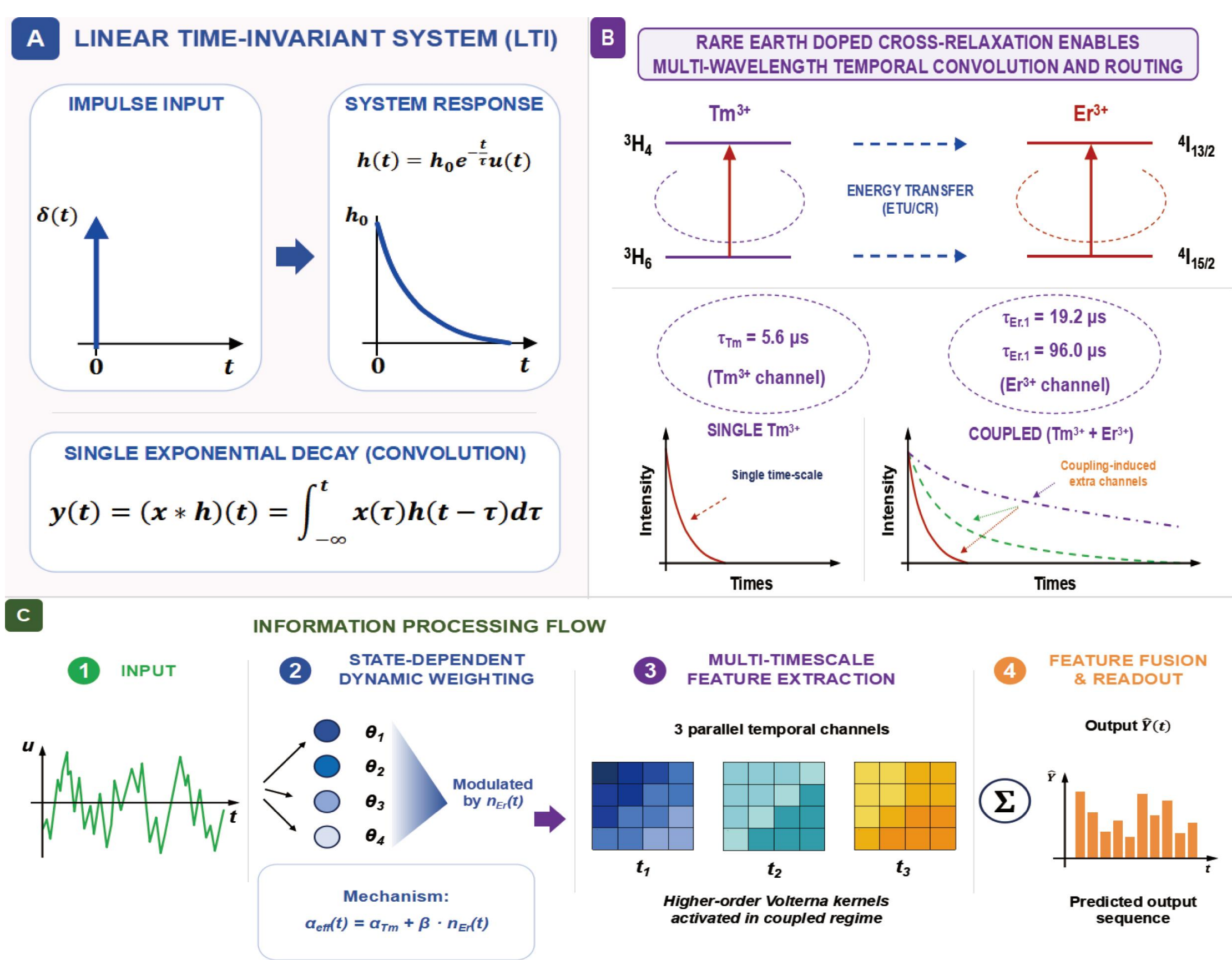

Figure 1. (a) Linear time-invariant optical device and its response function; (b) Schematic diagram of the rare-earth-doped ion system and its response function; (c) Schematic diagram of the optical information processing workflow realized using the rare-earth-doped ion system.

For most optical devices, the transfer function is linear and time-invariant, as shown in Figure 1a, which presents a fundamental limitation when handling recurrent neural network tasks that are time-dependent and nonlinear in nature. Constructing a nonlinear time-varying optical system is a necessary path to overcome this limitation, which requires the introduction of tunable nonlinear effects and dynamic memory mechanisms at the hardware level. To this end, we select $Tm^{3+}$@$Er^{3+}$ multilayer core-shell nanoparticles as the core physical platform to validate the computational potential of this rare-earth co-doped luminescent system. As illustrated in Figure 1b, this material system possesses multiple energy levels within its interior, naturally exhibiting multi-wavelength emission characteristics. The strong cross-relaxation and energy coupling processes among ionic energy levels endow its luminescence with nonlinear features, and each wavelength channel naturally corresponds to different time delays and memory windows. In the context of information processing, this means that a single input information stream is simultaneously subjected to multiple parallel temporal convolutions within the nanoparticle, with different wavelength channels serving as physical filters possessing distinct temporal receptive fields, extracting short-term, mid-term, and long-term historical features from the same time-series signal in a multi-dimensional manner. Crucially, the dynamical relaxation processes in rare-earth-doped ions are influenced by multiple input parameters, such as pump power, pulse width, and pulse interval. This means that the system's response function is dynamically modulated by the input signal sequence, as shown in Figure 1c. For a complex time-series signal, by decomposing it into degrees of freedom including power amplitude, signal duration, and signal interval, the device's dynamic nonlinear response can spontaneously allocate them to different relaxation channels, thereby constructing a flexible and dynamic nonlinear time-varying photonic information processing system. This purely physical, hardware-level adaptive gating mechanism further enhances the representation efficiency of phase-space states.

To translate the above dynamical mechanisms into macroscopic physical computing capability, we experimentally synthesized $Tm^{3+}$@$Er^{3+}$ multilayer core-shell upconversion nanoparticles (UCNPs) via a coprecipitation method(see Supplementary Materials for details), with the specific shell structure being $NaYF_4$:Tm@$NaYF_4$:Er. After spin-coating the nanoparticles into a thin film, we constructed an optical experimental test system for characterization, in which the time-series electrical signal carrying information is generated by an arbitrary waveform generator, amplified by a radio-frequency amplifier, and then used to drive an electro-optic modulator to perform intensity modulation on the pump light output from a continuous-wave laser. The modulated signal light is injected into the rare-earth-ion thin film, exciting nonlinear dynamical evolution. The resulting multi-channel fluorescence output is collected by an objective lens, spectrally separated by corresponding bandpass filters, converted into electrical signals by an avalanche photodiode, and acquired by an oscilloscope for readout (see Supplementary Materials for details).

Figure 2a presents the luminescence characterization of $Tm^{3+}$@$Er^{3+}$ multilayer core-shell nanoparticle thin films ($Er^{3+}$ doping concentration: 5%). In this architecture, $Tm^{3+}$ acts as the sensitizer, absorbing 1064 nm pump photons via ground-state absorption (${}^3H_6 \rightarrow {}^3H_5$) and excited-state absorption (${}^3F_4 \rightarrow {}^3F_2$) and emitting near 800 nm (${}^3H_4 \rightarrow {}^3H_6$), while transferring excited-state energy to $Er^{3+}$ through energy transfer upconversion (ETU) and cross-relaxation (CR). The activated $Er^{3+}$ subsequently produces characteristic emission at 550 nm (${}^2H_{11/2}/{}^4S_{3/2} \rightarrow {}^4I_{15/2}$) and 650 nm (${}^4F_{9/2} \rightarrow {}^4I_{15/2}$). Consequently, a single 1064 nm pump is split into three spectrally distinct emission channels within one nanoparticle, establishing a multi-wavelength parallel computational architecture without external demultiplexing.

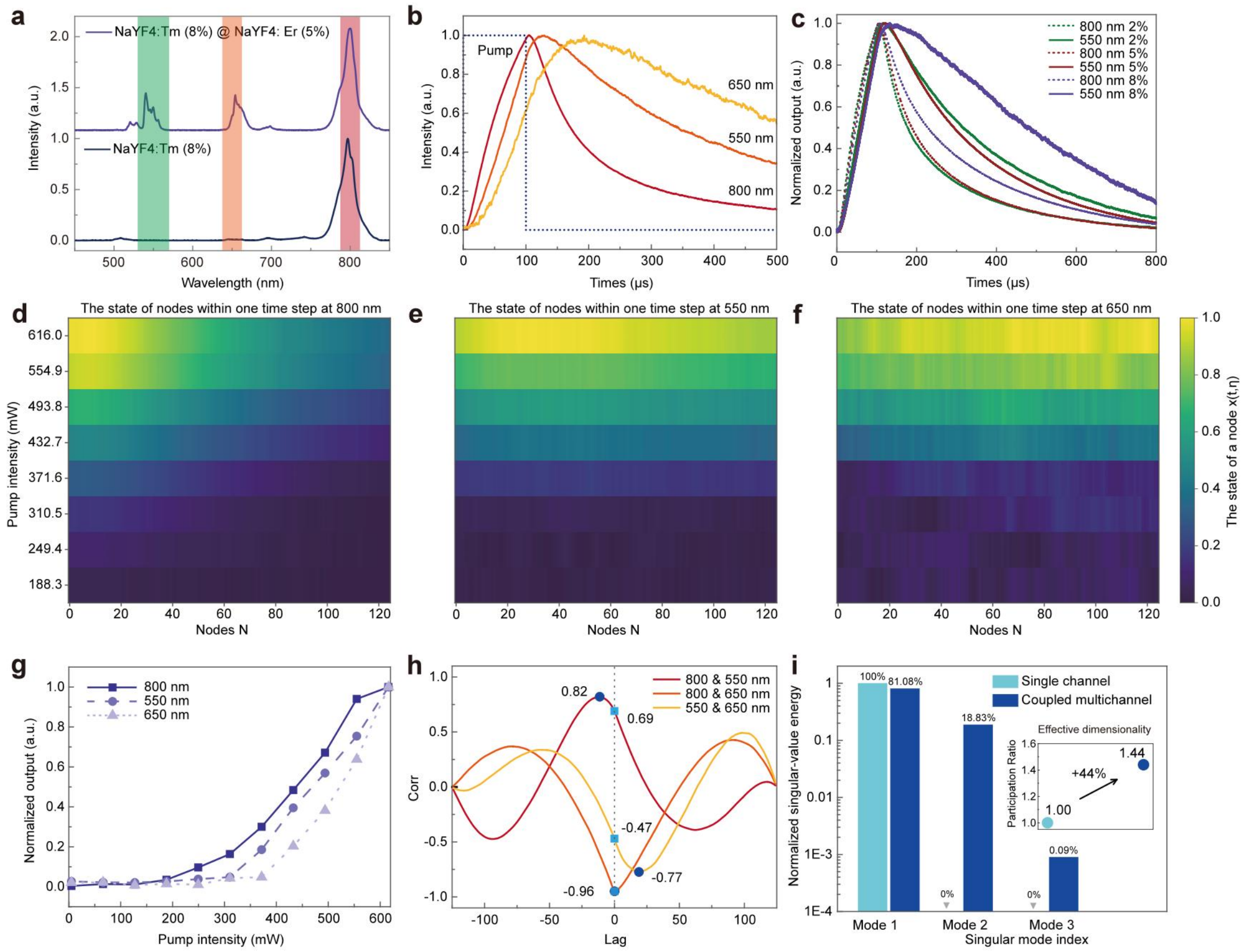

Figure 2. Characterization of dynamic luminescence properties of the $Tm^{3+}$@$Er^{3+}$ core-shell nanoparticles. (a) Emission spectra of $Tm^{3+}$-only and $Tm^{3+}$@$Er^{3+}$ samples under 1064 nm excitation. (b) Decay dynamics of the three emission channels (800 nm, 550 nm, and 650 nm) under identical pumping conditions, with decay times of 5.6 μs, 19.2 μs, and 96.0 μs, respectively. (c) Evolution of luminescence lifetimes in the 800 nm and 550 nm channels as a function of $Er^{3+}$ doping concentration (2%–8%). (d-f) Instantaneous normalized luminescence intensity of the 800 nm, 550 nm, and 650 nm channels as a function of time and pump power over one cycle following the pump pulse. (g) Peak luminescence response of each emission band as a function of input pump power. (h) Normalized cross-correlation analysis among the three wavelength channels. (i) Singular value spectra and participation ratios (PR) of the single-channel and coupled multi-channel systems.

These three channels are not equivalent: differences in energy-transfer pathways and phonon-assisted relaxation rates impose markedly heterogeneous decay dynamics. Under identical pumping conditions (Figure 2b), the 800 nm, 550 nm, and 650 nm channels exhibit decay times of 5.6 μs, 19.2 μs, and 96.0 μs, respectively—a 17-fold span across a single excitation cycle. Each channel therefore corresponds to a distinct memory window, collectively expanding the system's temporal receptive field across short, intermediate, and long time scales from a single input stream. The coupling between $Tm^{3+}$ and $Er^{3+}$ further breaks time-invariance. As $Er^{3+}$ doping concentration increases from 2% to 8% (Figure 2c), the 800 nm lifetime monotonically shortens while the 550 nm lifetime first decreases then rises—a non-monotonic signature of competing CR and ETU pathways that directly modulates the system's temporal transfer function. To characterize the time-varying response, we scanned pump power across one excitation cycle (Figures 2d–f). When the input signal power is relatively low (<350 mW), the system's luminescence response is predominantly concentrated in the 800 nm channel and its corresponding short-lifetime temporal response. However, when the input power exceeds a specific threshold (>350 mW), the energy transfer channels in the co-doped system are activated; in addition to the fast response of the 800 nm channel itself, energy is also transferred to the longer-delay 550 nm and 650 nm channels. To more intuitively distinguish these differences, we extracted the peak luminescence response of each band and plotted their corresponding mapping relationships with the input pump power, as shown in Figure 2g. It can be observed that the response curves of each luminescence channel exhibit significant differences in threshold power and gain slope, demonstrating strongly nonlinear mapping characteristics.

To quantify the representational advantage of this multi-channel architecture, we constructed a state matrix **H** ∈ ℝ^{N×3} by sampling the output of each channel at $N$ time nodes within one time step:

$$\mathbf{H}=[\boldsymbol{h}^{\lambda_1}(t) \quad \boldsymbol{h}^{\lambda_2}(t) \quad \boldsymbol{h}^{\lambda_3}(t)] \tag{1}$$

where $\boldsymbol{H}\in\boldsymbol{R}^{N\times 3}$, $t=0,1,\ldots,N-1$, $\lambda_1$, $\lambda_2$, $\lambda_3$ denote the three wavelength channels. Each row of the matrix corresponds to the outputs of the three wavelengths at the same time instant, while each column corresponds to the complete response of a single wavelength channel over the N time nodes. As shown in Figure 2h, normalized cross-correlation analysis reveals three distinct inter-channel relationships: Channel 1(800 nm) and channel 2(550 nm) share similar waveform morphology but are offset by a lag of −12 time steps (peak correlation 0.818), providing delayed replicas that aid historical-feature reconstruction; Channels 1 and 3(650 nm) exhibit strong anti-phase correlation (peak −0.772 at lag 19), adding discriminative dimensions at the readout stage; Channels 2 and 3 are nearly synchronous yet strongly anti-correlated at zero lag (−0.958), encoding complementary simultaneous dynamics. Besides, singular value decomposition (SVD) of **H** yields a participation ratio (PR) of 1.44, a 44% gain over the single-channel baseline (PR = 1.00), confirming that the coupled system occupies a substantially higher-dimensional subspace of the state space (Figure 2i; see Supplementary Information for the full procedure). This expanded dimensionality enables input signals to be projected onto a richer set of linearly independent basis functions, directly improving the system's capacity for complex time-series feature extraction.

To quantify how these physical degrees of freedom translate into computational capacity, we evaluated linear and nonlinear memory depths using the standard short-term memory (STM) capacity test (Supplementary Note X). As shown in Figure 3a, we used a uniform random sequence u(t) ~ U(0,1) of length 2000 as the encoding input, and measured and computed the luminescence responses of the single-ion $Tm^{3+}$ system, the $Tm^{3+}$@$Er^{3+}$ single-channel system, and the $Tm^{3+}$@$Er^{3+}$ three-channel system. Figure 3b presents the decay curves of the reconstruction accuracy of past input information as a function of delay step k. It can be observed that for the single $Tm^{3+}$ system, $MC_k$ at the 800 nm emission band decays rapidly to near zero after k > 2, with its effective memory depth strictly limited by the intrinsic relaxation time and a memory capacity of only MC = 0.7. In contrast, for the $Tm^{3+}$@$Er^{3+}$ co-doped system, the $MC_k$ of the single 800 nm channel (red curve) increases significantly, with MC rising to 1.4, indicating that the cross-relaxation process effectively extends the memory time. After incorporating all three wavelength channels (800, 550, and 650 nm) in parallel into the computation, it can be seen that not only does the system maintain a significantly higher capacity than the single-channel case at short-to-medium delays (k = 2–4), but it also contributes non-negligible memory components at long delays (k = 4–10), with the total memory capacity increasing to MC = 3.1.

To resolve the mechanistic origin of this gain, we decomposed the total capacity via Volterra series expansion into linear, nonlinear (diagonal), and cross-memory (off-diagonal) components (Figure 3c). In the single $Tm^{3+}$ system, the linear capacity dominates (0.6), the quadratic nonlinear capacity is ~0.05, and the cross capacity is zero, which consistent with predominantly linear single-ion relaxation dynamics. Upon forming the coupled $Tm^{3+}$@$Er^{3+}$ system, the linear capacity increases substantially due to extended effective memory time, but the critical change is the emergence of nonzero cross-memory capacity. Cross terms encode the joint state of the input at two distinct historical instants, a capability that cannot arise from any superposition of independent single-time-instant responses. Their presence is therefore a definitive indicator that the system has acquired genuine nonlinear dynamical memory. In the Supplementary Materials, we further provide rigorous mathematical derivations proving that the inter-ion energy coupling can directly modulate the magnitude of the cross capacity.

A natural subsequent question is whether there exists an optimal doping concentration or coupling strength that maximizes the system's memory capacity or optimizes its computational capability. To address this, we systematically varied the $Er^{3+}$ doping concentration in the shell layer of the co-doped nanoparticles (from 0% to 8%) and measured the memory capacity of each system under fixed pumping conditions. As shown in Figure 3d, MC rises sharply from 1.49 (0%) to 2.6 at 2% (+75%), peaks at MC = 3.1 at 5%, then declines to 2.88 at 8%. The physical mechanism is a competition between two opposing effects of coupling strength on system dynamics. In the low-doping regime (<5%), the coupling-induced additional decay term $\beta_{Er}$ remains smaller than the intrinsic $Tm^{3+}$ decay rate $\alpha_{Tm}$; cross-relaxation activates higher-order Volterra kernels (Eqs. S8–S9), generating cross-memory capacity while moderately extending the effective memory time. At 5%, $\beta_{Er} \approx \alpha_{Tm}$ and the spectral radius approaches the critical value $\rho=1$, the edge-of-stability condition that simultaneously maximises both memory depth and nonlinear expressivity. Beyond 5%, $\beta_{Er} > \alpha_{Tm}$ drives the effective decay rate into an overdamped regime: the spectral radius collapses well below unity, historical information decays too rapidly to be retained across successive input steps, and MC falls accordingly.

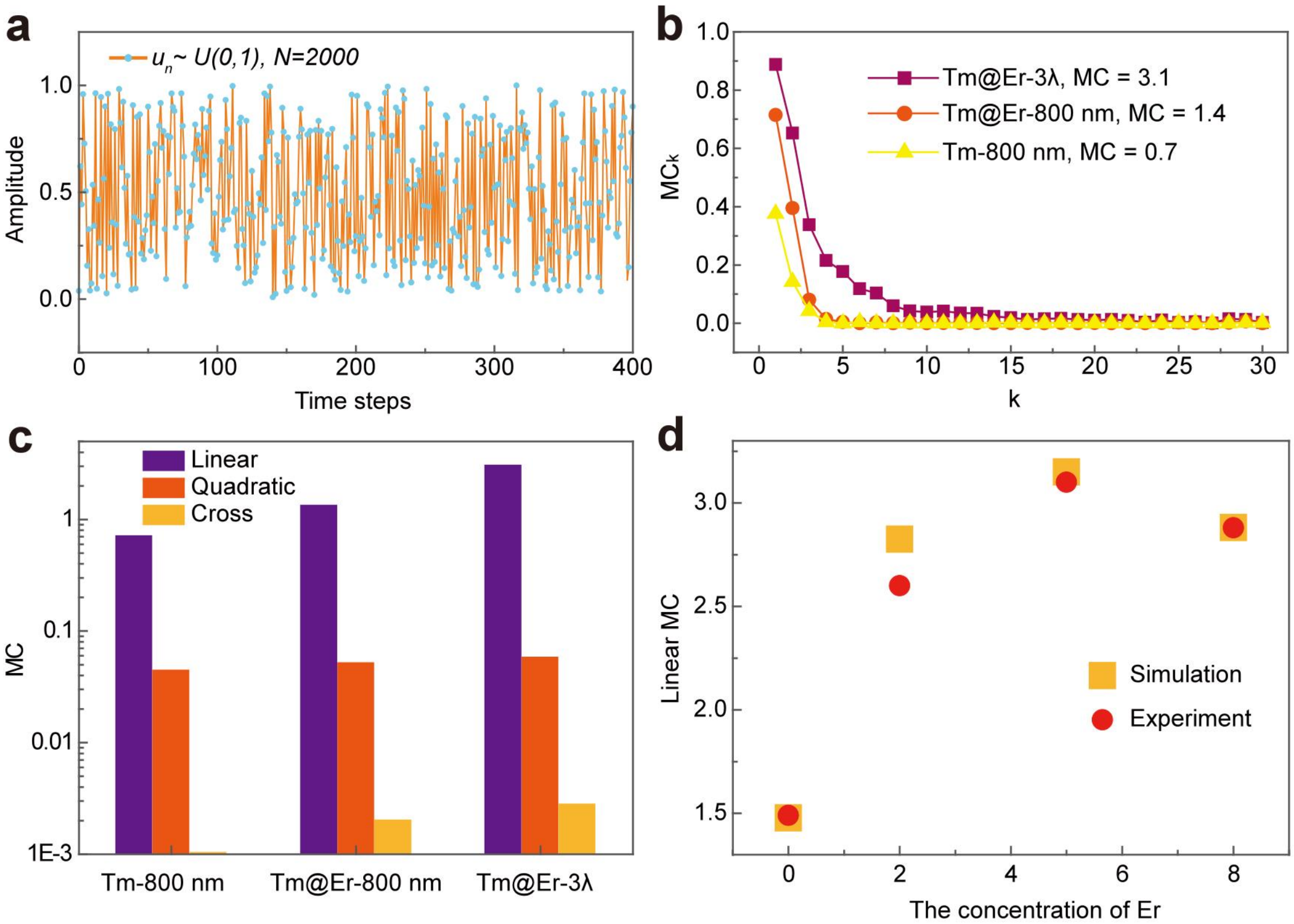

Figure 3. Memory capacity characterization of the $Tm^{3+}$@$Er^{3+}$ system. (a) A segment of the 2000-point uniform random input sequence u(t) used for encoding. (b) Reconstruction accuracy $MC_k$ versus delay step k for the single $Tm^{3+}$ system, $Tm^{3+}$@$Er^{3+}$ single-channel, and three-channel configurations. (c) Decomposition of total memory capacity into linear, diagonal nonlinear, and cross components for each system. (d) Memory capacity as a function of $Er^{3+}$ doping concentration (0%–8%)

Through the systematic physical characterization of the above doped-coupled ion system, we have verified the potential of the state-dependent time-varying properties and multi-channel responses inherent to rare-earth-doped ion systems in enhancing computational performance. Building on this, we employed the rare-earth ion-coupled system as a reservoir for time-series prediction tasks. The reservoir computing architecture is characterized by fixed connection weights in the hidden layer, with only the output layer being trained, which effectively isolates interference from training algorithms and network structure optimization, allowing for a direct assessment of the computational capability inherent to the nonlinear physical system itself.

In the experiments, we first encoded the time series $u(t) \in R^{L\times 1}$ into an intensity-modulated 1064 nm pump signal with a time step of T = 100 μs per input point, and fed it into the rare-earth ion-coupled reservoir. Within the reservoir computing framework, virtual nodes were constructed using a time-multiplexing strategy: during each input period T, the oscilloscope continuously sampled the fluorescence response of the rare-earth thin film at a fixed time interval θ = T/N, evenly dividing the temporal response of a single pulse into N time nodes, with each sampling point corresponding to one virtual node $x_\sigma(t)$ (σ = 1, 2, …, N). In this experiment, N was set to 125, corresponding to a node interval of θ = 0.8 μs. With the clocks of the arbitrary waveform generator and the oscilloscope synchronized, the output signals were read out from three wavelength bands (800 nm, 550 nm, and 650 nm) and combined into a state matrix $[x_{\lambda_1}(t) \quad x_{\lambda_2}(t) \quad x_{\lambda_3}(t)] \in R^{L\times 3N}$, where N is the number of virtual nodes. Ridge regression was then employed to train the readout layer weights $W_{out} \in R^{3N\times 1}$, with the training target set to $\hat{y}(t)=u(t+1)$. Prediction performance was quantitatively evaluated using the normalized mean squared error (NMSE)( detailed definitions provided in the Supplementary Materials).

We selected the classic Mackey-Glass (MG) chaotic sequence as a benchmark task, with the goal of single-step prediction. The MG equation is a type of nonlinear delay differential equation. The MG sequence exhibits both strong nonlinearity and long-term delay dependence, making it capable of simultaneously evaluating the nonlinear mapping capability and memory capacity of neural networks. The parameters were set to $\alpha$=0.2、$\gamma$=0.1、$\beta$=10、$\tau$=17, under which the system exhibits pronounced chaotic behavior. In the experiments, the sequence length L was set to 2000; the first and last 100 data points were discarded to eliminate transient effects and boundary crosstalk, and the remaining data were divided into a training set (the first 1200 points) and a test set (the last 600 points).

Figures 4(b) and 4(c) present the prediction results on the MG task for the single $Tm^{3+}$ ion reservoir and the $Tm^{3+}$@$Er^{3+}$ co-doped reservoir, respectively. It can be observed that the predicted trajectory of the single-ion system deviates slightly from the ground truth, reflecting that its limited memory depth and nonlinear mapping capability are insufficient to perfectly capture the long-term dependence structure of the MG sequence. In contrast, the three-channel coupled output of the co-doped system significantly suppresses error accumulation, with the predicted trajectory closely matching the ground truth over the entire test interval, demonstrating the advantage of multi-channel, multi-scale responses in complex temporal prediction tasks. Figure 4(d) provides a direct comparison of the MG prediction performance (NMSE) between the single-ion and co-doped reservoirs as a function of the number of virtual nodes. At N = 125, the single-ion reservoir achieves an NMSE of 0.008, while the three-channel co-doped coupled system reduces the NMSE to $1.2 \times 10^{-3}$, representing a nearly sevenfold improvement. This result further confirms that the energy transfer and cross-relaxation processes between $Tm^{3+}$ and $Er^{3+}$ in the co-doped system enable the reservoir to more adequately capture the nonlinear structures and long-range correlations inherent to the chaotic sequence.

To further validate the generalization capability of the reservoir across different chaotic regimes, we applied the same physical system to the Santa Fe chaotic laser time-series prediction task. The Santa Fe dataset consists of intensity measurements from a far-infrared laser operating in a chaotic state, whose dynamics are described by a Lorenz-like system, producing output sequences characterized by sharp and irregular spike events. These features impose high demands on the nonlinear response and fast transient capture capability of the prediction model. Figure 4(e) shows the prediction results of the three-channel coupled reservoir on the Santa Fe test set. The predicted sequence (red curve) successfully reproduces both the slowly varying envelope modulation and the fast spike events of the ground truth (black curve), achieving an NMSE of 0.021, indicating that the reservoir is capable of simultaneously handling multi-timescale dynamics.

Figure 4(f) compares the performance of our system with that of recently reported state-of-the-art reservoir computing systems for time-series prediction[25-31]. On the Mackey-Glass task, the rare-earth ion-coupled reservoir achieves an NMSE of $1.2 \times 10^{-3}$ with 125 virtual nodes; on the Santa Fe dataset, it achieves an NMSE of $2.1 \times 10^{-2}$ under the same configuration. These metrics surpass most existing methods with a more compact node scale, placing the core performance at the advanced level among recent studies. These results demonstrate that the rare-earth ion-coupled reservoir achieves excellent prediction accuracy on both benchmark tasks, validating the great potential of rare-earth-doped materials for constructing high-efficiency, compact physical computing hardware.

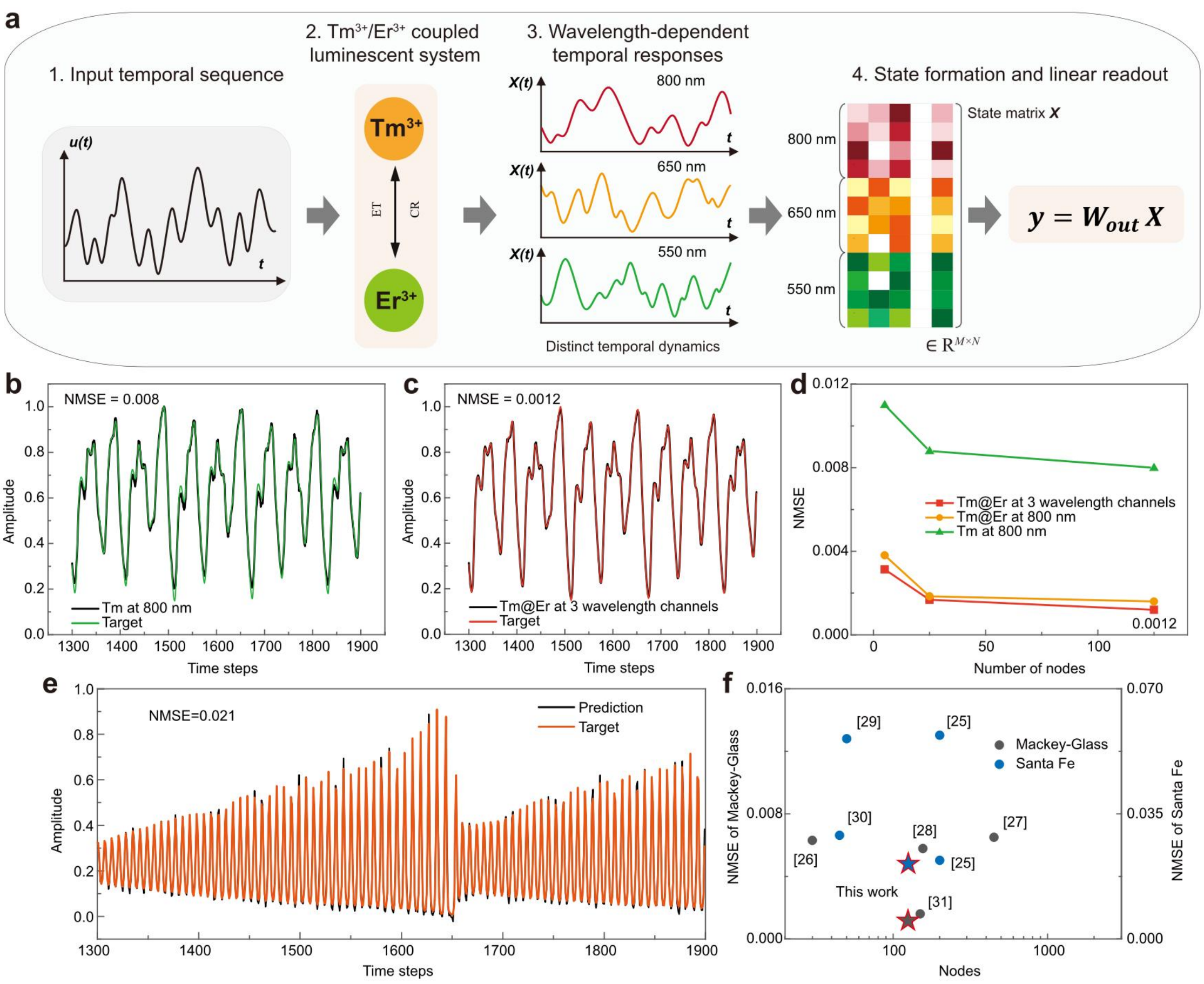

Figure 4. Chaotic time-series prediction using the rare-earth ion-coupled reservoir. (a) Schematic illustration of the multi-wavelength rare-earth ion-coupled reservoir computing system. (b) Prediction results of the single $Tm^{3+}$ reservoir on the Mackey-Glass task. (c) Prediction results of the $Tm^{3+}$@$Er^{3+}$ co-doped reservoir on the MG task in the time domain. (d) Comparison of NMSE performance between the single-ion and co-doped reservoirs as a function of the number of virtual nodes on the MG task. (e) Prediction results of the co-doped reservoir on the Santa Fe chaotic laser time-series task. (f) Performance comparison between our system and state-of-the-art reservoir computing systems on both benchmark tasks.

## Discussion and Conclusion

For a long time, the standard paradigm in the field for achieving high-dimensional information processing capacity within resource-constrained microscopic nodes has been to "seek outward", relying on bulky macroscopic delay lines or complex external feedback circuits for temporal multiplexing. This work breaks through this limitation and innovatively demonstrates the feasibility of "mining inward": by directly exploiting the many-body interactions inherent to rare-earth materials, we construct a novel physical computing paradigm with time-varying characteristics. The intrinsic dynamical parameters of rare-earth materials, such as relaxation rates and coupling strengths, can respond to and depend on the system's own state in a real-time, nonlinear manner, thereby dynamically modulating their information processing capability. This endogenous higher-order temporal feature not only fundamentally eliminates the reliance on external physical delay lines, but also provides a solution for high-density edge physical AI hardware that does not require "trading space for time." This discovery not only introduces a novel intrinsic computing paradigm to the field of physical computing, but also reveals the broad prospects of deeply mining the "inward evolution" potential of physical systems to overcome traditional information processing bottlenecks. The time-varying mechanism of rare-earth materials is of general relevance, with its core lying in the nonlinear dynamic response of internal system parameters to the system state. Therefore, other physical systems can follow the same validation pathway: first, examine their capability for time-varying dynamic parameter modulation, that is, whether their intrinsic dynamical parameters (e.g., refractive index, electrical resistance, magnetization) can undergo nonlinear dynamic changes in response to system states (e.g., electric field, magnetic field, optical intensity, temperature); and second, evaluate the resulting improvements in computational capability (such as significant enhancements in memory capacity and nonlinear mapping capacity) to demonstrate their computational potential [7, 20, 23-24]. For example, in silicon-based integrated photonics, state-dependent dynamic modulation of microring resonator Q-factors or refractive indices can be achieved through electro-optic or thermo-optic effects.

In summary, this study provides new theoretical pathways and design principles for constructing next-generation high-efficiency neuromorphic computing hardware. With the transition from benchtop optical systems to chip-integrated platforms, the rare-earth-doped architecture promises to become a compact, real-time inferencing physical computing solution that, by harnessing the inherent dynamical complexity of the physical world, achieves efficient intelligent computing beyond the von Neumann architecture.

## Methods

**Material Synthesis.** Reagents: Anhydrous erbium chloride (99.9%), yttrium chloride hexahydrate (99.9%), anhydrous thulium chloride (99.9%), trifluoroacetic acid, erbium oxide, yttrium oxide, sodium hydroxide, ammonium fluoride, 1-octadecene, and oleic acid were used. All reagents were purchased from Aladdin and used without further purification. In this study, upconversion nanoparticles (UCNPs) with a $NaYF_4$:Tm@$NaYF_4$:Er multilayer core-shell structure was synthesized via a coprecipitation method.

Synthesis of $NaYF_4$:Tm (8%) core: $YCl_3 \cdot 6H_2O$ (0.92 mmol) and $TmCl_3$ (0.08 mmol) were added to a 100 mL three-neck flask containing 6 mL of oleic acid (OA) and 15 mL of 1-octadecene (ODE). The mixture was stirred and heated to 160 °C under argon and maintained for 30 min to form lanthanide oleate complexes. Here, oleic acid serves as a ligand to coordinate with $Ln^{3+}$ for regulating the reaction rate, while 1-octadecene acts as a high-boiling-point solvent to provide a stable reaction environment. After the system was cooled to 50 °C, 10 mL of a methanol solution containing NaOH (2.5 mmol) and $NH_4F$ (4 mmol) was slowly added dropwise into the flask and stirred for 30 min. The solution was heated to 100 °C and maintained for 30 min to evaporate the methanol, then further heated to 300 °C and held for 1 h. The final product was washed by centrifugation with ethanol, purified by dispersion in cyclohexane, yielding pure $NaYF_4$:Tm core nanocrystals.

Shell coating of $NaYF_4$:Er (5%): First, rare-earth trifluoroacetate precursors were prepared. Based on the molar amount of $NaYF_4$:Er (5%) for crystal growth and the ion doping ratio, the corresponding amounts of $Y_2O_3$ and $Er_2O_3$ were weighed into a 100 mL three-neck flask. Subsequently, 5 mL of trifluoroacetic acid ($CF_3COOH$) and 5 mL of distilled water were added, and the mixture was heated to 85 °C under reflux. After the solution became clear, the side port was opened and argon was introduced to remove excess trifluoroacetic acid and water, yielding the rare-earth trifluoroacetate salts. Then, 2 mmol of sodium trifluoroacetate ($CF_3COONa$), 10 mL of oleic acid (OA), 10 mL of 1-octadecene (ODE), and 5 mL of the core nanocrystal dispersion in cyclohexane were sequentially measured and added into the flask. Under an argon atmosphere, the reaction solution was heated to 120 °C and maintained for 30 min to remove the added cyclohexane, then heated to 300 °C for crystal growth for 48 min, and subsequently cooled to room temperature. The resulting reaction solution was precipitated with anhydrous ethanol, then centrifuged (8000 rpm, 5 min). The supernatant was discarded to obtain a white precipitate, which was then redispersed in cyclohexane by ultrasonication and washed with ethanol (1:2 volume ratio) by centrifugation twice; each centrifugation was followed by 5 min of ultrasonication to ensure thorough washing. The shell thickness could be controlled by adjusting the amount of trifluoroacetate precursors added. The final product was dispersed in cyclohexane after centrifugal washing.

**Optical Measurements.** A 1064 nm continuous-wave laser (CW, MIL-S-1064-SM, CNI) was coupled through a single-mode fiber and an aspheric collimating lens. After modulation by an electro-optic modulator (EOM, EO-AM-R-20-C1, Thorlabs), the pump light carrying information was directed onto the rare-earth-ion thin film. The output fluorescence was collected by an objective lens, passed through bandpass filters of different wavelength bands, and detected by an avalanche photodiode (APD, 2051-FS-M, Newport) and an oscilloscope (MSO3054X, UNI-T). Signal modulation was accomplished by an arbitrary waveform generator (AWG, UTG9604T, UNI-T) and a radio-frequency amplifier, with a photodetector used to monitor the modulated signal. The oscilloscope sampling length was set to 250 kSa, and 64-times averaging was employed during fluorescence signal acquisition. The bandpass filters used were centered at 800 nm (FWHM = 25 nm, MBF10-800-25, LBTEK), 550 nm (FWHM = 40 nm, OFBH140-550, JCOPTIX), and 650 nm (FWHM = 25 nm, MBF10-650-25, LBTEK).

## Acknowledgement

The authors acknowledge support by National Key Research and Development Program of China (Grant Nos. 2024YFB2809200, 2022YFB3505700), National Natural Science Foundation of China (Grant Nos. 6233000076, 12334016, 11934012, 12025402, 12261131500, 92250302, 12474376 and 62305084), Guangdong Basic and Applied Basic Research Foundation (2023A1515011746, 2024B1515020060, 2023B1212010003, 2022A1515012108), Shenzhen Fundamental Research Projects (Grant Nos. GXWD 2022081714551), Guangdong Provincial Quantum Science Strategic Initiative(GDZX2406002, GDZX2306002), Shenzhen Science and Technology Program (Grant Nos. JCYJ20230807094401004), Shenzhen Fundamental research project (JCYJ20241202123719025, JCYJ20210324120402006, JCYJ20200109112805990, GXWD 20220817145518001, RCYX20231211090432059), the Fundamental Research Funds for the Central Universities (Grant No. HIT.OCEF.2024020, 2022FRFK01013). Guangdong and Hong Kong Universities '1+1+1' Joint Research Collaboration Scheme.

## Data availability

The data that support the findings in this study is available from the corresponding authors upon reasonable request.

## Competing interests

The authors declare no competing interests.

## Author contributions

Q.S., C. H., L.J., conceived the idea and supervised the research. J.C., X.L., and L.J. prepared the experimental materials. J.C., W.S., S.G., performed the experimental measurements. J.F., and J.C. did the simulation. Q.S., C.H., A.D. and J.Y. analyzed the results. C.H. prepared the manuscript from all authors' contributions.